\documentclass[aps,prl,floatfix,twocolumn,superscriptaddress,twoside]{revtex4-1}
\usepackage{amsmath,amssymb,amsfonts}
\usepackage{graphicx}
\usepackage[english]{babel}
\usepackage{psfrag}
\usepackage{color}
\usepackage{tabularx}
\usepackage{times}
\usepackage{hyperref}
\usepackage{float}

\begin{document}

\title{Quantum Criticality and Spin Liquid Phase in the Shastry-Sutherland model}

\author{Jianwei Yang}
\affiliation{Beijing Computational Science Research Center, 10 East Xibeiwang Road, Beijing 100193, China}

\author{Anders W. Sandvik}
\email{sandvik@bu.edu}
\affiliation{Department of Physics, Boston University, 590 Commonwealth Avenue, Boston, Massachusetts 02215, USA}
\affiliation{Beijing National Laboratory for Condensed Matter Physics and Institute of Physics, Chinese Academy of Sciences, Beijing 100190, China}

\author{Ling Wang}
\email{lingwangqs@zju.edu.cn}
\affiliation{Department of Physics, Zhejiang University, Hangzhou 310000, China}

\begin{abstract}
Using the density-matrix renormalization group method for the ground state and excitations of the Shastry-Sutherland
spin model, we demonstrate the existence of a narrow quantum spin liquid phase between the previously known plaquette-singlet 
and antiferromagnetic states. Our conclusions are based on finite-size scaling of excited level
crossings and order parameters. Together with previous results on candidate models for deconfined quantum criticality and 
spin liquid phases, our results point to a unified quantum phase diagram where the deconfined quantum-critical point separates a line 
of first-order transitions and a gapless spin liquid phase. The frustrated Shastry-Sutherland model is close to the critical point 
but slightly inside the spin liquid phase, while previously studied unfrustrated models cross the first-order line. We also argue 
that recent heat capacity measurements in SrCu$_2$(BO$_3$)$_2$ show evidence of the proposed spin liquid at pressures between 2.6 and 3 GPa.
\end{abstract}

\date{\today}

\maketitle

The quasi two-dimensional (2D) $S=1/2$ quantum magnet SrCu$_2$(BO$_3$)$_2$ \cite{Kageyama99,Miyahara99,koga00} has emerged 
\cite{Zayed17,Zhao19,Lee19,Bettler20,Guo20,Jimenez20} as the most promising realization of a deconfined quantum-critical point (DQCP) 
\cite{Senthil04,Sachdev08,Sandvik07},  where a state spontaneously forming a singlet pattern meets an antiferromagnetic 
(AF) state in a phase transition associated with fractionalized excitations (spinons). The intralayer interactions of 
the Cu spins correspond to the Shastry-Sutherland (SS) model \cite{Shastry81}, with highly frustrated AF interdimer ($J$)
and intradimer ($J$') Heisenberg couplings. The SS model has three known ground states
versus $g=J/J'$; a dimer singlet (DS) state for small $g$ \cite{Shastry81}, a N\'eel AF state for large $g$,
and a two-fold degenerate plaquette-singlet (PS) state for $g \in [0.68,0.77]$ \cite{koga00,Corboz13,Nakano18,Lee19}. 

At ambient pressure SrCu$_2$(BO$_3$)$_2$ is in the DS phase \cite{Kageyama99,Miyahara99} but the other SS phases have been 
anticipated under high pressure \cite{Waki07}. Recent heat capacity \cite{Guo20,Jimenez20}, neutron scattering \cite{Zayed17},
and Raman \cite{Bettler20} experiments have indeed confirmed some variant \cite{Boos19,Shi21} of the PS phase (from 1.7 to 2.5 GPa at 
temperatures $T <2$ K) and an AF phase (between 3 and 4 GPa below 4 K). A direct PS--AF transition may then be expected between 
2.6 and 3 GPa \cite{Sun21} at temperatures not yet reached.

Here we show that the above picture is incomplete. Using the density-matrix renormalization group (DMRG) method \cite{White92},
we study the ground state and low-lying excitations of the SS model. Based on the lattice-size dependence of the level spectrum
and order parameters, we conclude that a narrow gapless spin liquid (SL) phase intervenes between the PS and AF phases. In light 
of this finding, the absense of signs of any phase transition between 2.6 and 3 GPA \cite{Guo20,Jimenez20} opens the intriguing 
prospect of an SL phase in SrCu$_2$(BO$_3$)$_2$.

\begin{figure}[b]
\includegraphics[width=40mm]{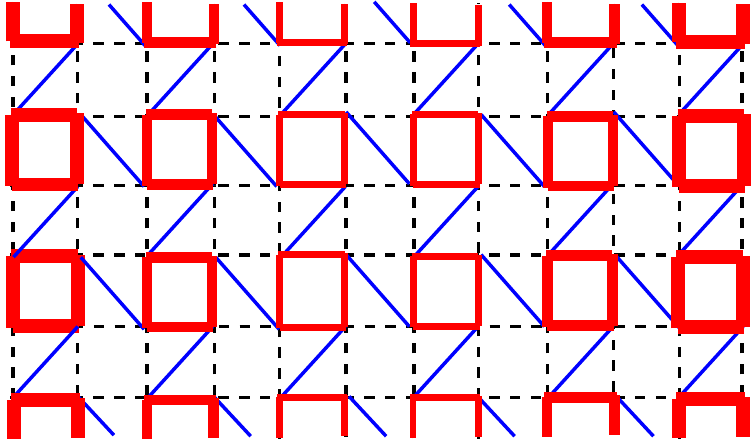}
\vskip-1mm
\caption{The SS lattice with open $x$ and periodic $y$ boundary conditions. The lengths $L_x$ and $L_y$ are both even.
Nearest neighbors are coupled at strength $J$ by Eq.~(\ref{ham}) and the blue diagonal links represent the dimer couplings $J'$. The open edges
break the $\mathbb{Z}_2$ symmetry of the PS phase, thus inducing a singlet density pattern as indicated schematically by the thickness
of the red lines.}
\label{lattice}
\end{figure}

{\it DMRG calculations.}---The SS model with AF couplings $J$ between first neighbor spins $\left<ij\right>$ and $J'$ on a subset of
second neighbors $\left<ij\right>'$ is illustrated in Fig.~\ref{lattice}. The Hamiltonian is \cite{Shastry81}
\begin{equation}
H=J\sum_{\left<ij\right>} \mathbf{S}_i \cdot \mathbf{S}_j + J'\sum_{\left<ij\right>'} \mathbf{S}_i \cdot \mathbf{S}_j,
\label{ham}
\end{equation}
here on $L_x \times L_y$ cylinders \cite{Schollwock11,Stoudenmire13} with open and periodic boundaries in the $x$ and $y$ direction,
respectively, and $L\equiv L_y =2n$, $L_x=2L$. In this geometry, the model has a preferred singlet pattern which minimizes the boundary energy in the PS
phase; thus the two-fold degeneracy is broken and the ground state is unique, as illustrated in Fig.~\ref{lattice}.

We have developed efficient procedures for calculating not only the ground state with full SU(2) symmetry \cite{Wbaum12,Gong14}, but also successively
generating excited states by orthogonalizing to previous states \cite{McCulloch07,Wang18,Lemm20}. Imposing stringent convergence criteria for given 
Schmidt number $m$, we have reached sufficiently large $m$ for reliably extrapolating to discarded weight $\epsilon_m = 0$ (Supplemental Material \cite{sm}) 
for $L$ up to $10$, $12$, or $14$ depending on quantity (energies and order parameters). 
Any remaining errors in the results are small on the scale of the graph symbols in the figures presented below.

\begin{figure}[t]
\includegraphics[width=70mm]{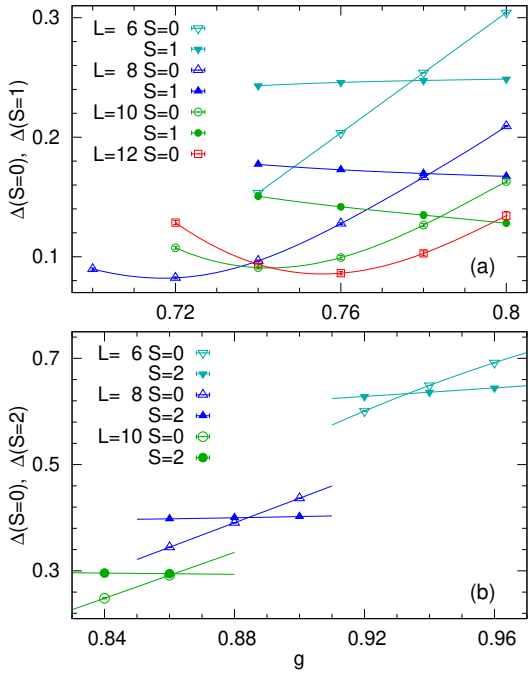}
\vskip-1mm
\caption{(a) The lowest singlet and triplet gaps vs $g$ in the neighborhood of the expected quantum phase transition out of the PS phase.
(b) The lowest singlet and quintuplet gaps for $g$ inside the AF phase, close to its quantum phase transition. The curves are polynomial
fits.}
\label{stgaps}
\end{figure}

We focus on the window $g \in [0.7,0.9]$, which straddles the PS and AF phases. The ground state of the system is always
a singlet, and we analyze the gaps $\Delta(S)$ to the lowest excited singlet ($S=0$), triplet ($S=1$), and qintuplet ($S=2$). Finite-size crossings
of excited levels with different spin are often used indicators of quantum phase transitions in spin chains \cite{Nomura92,Eggert96,Sandvik10,Suwa15},
and this method was also applied to the 2D $J$-$Q$ \cite{Suwa16} and $J_1$-$J_2$ \cite{Wang18,Nomura20,Ferrari20} Heisenberg models. We here use
level crossings to detect the transitions out of the PS phase and into the AF state, following Ref.~\onlinecite{Wang18}
closely. We also study the PS and AF order parameters to corroborate the quantum phases and phase transitions.

\begin{figure}[t]
\includegraphics[width=65mm]{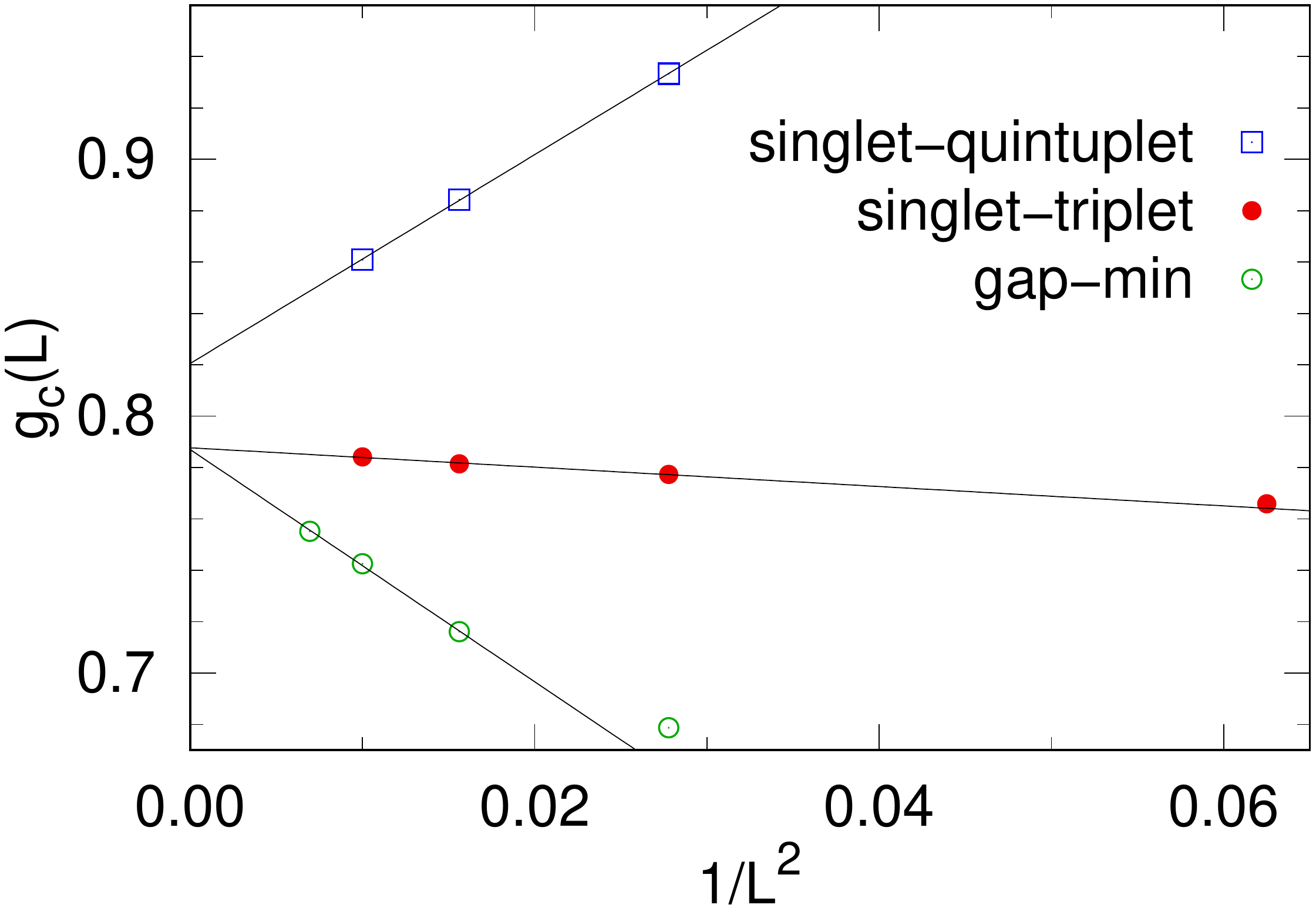}
\vskip-1mm
\caption{Locations of gap crossings and singlet minimums, with the lines showing linear-in-$1/L^2$ fits. The $L=4$ singlet-quintuplet point
is at $g \approx 1.1$, falling very close to the fitted line. The extrapolated critical points are $g_{c1} = 0.788 \pm 0.002$ and
$g_{c2} = 0.820 \pm 0.002$.}
\label{gcfits}
\end{figure}

We graph singlet and triplet gaps in Fig.~\ref{stgaps}(a) and similarly singlet and quintuplet gaps in Fig.~\ref{stgaps}(b), in $g$ windows
where gap crossings are observed. In Fig.~\ref{gcfits} we analyze the gap crossing points and the singlet minimum that is also observed in
Fig.~\ref{stgaps}(a).  Given the previous empirical observations of crossing-point drifts in 2D systems \cite{Suwa16,Wang18},
we graph the results versus $1/L^2$ and find almost perfect linear behaviors. Interesting, the singlet-triplet crossing and the singlet
minimum both extrapolate to $g_{c1} \approx 0.79$, while the singlet-quintuplet points scale to a higher value; $g_{c2} \approx 0.82$.

It was previously shown \cite{Sandvik10,Wang18} that the crossing point between the lowest singlet and quintuplet levels is a useful finite-size
estimator for a quantum phase transition into an AF phase, given that the lowest $S>0$ states are Anderson quantum rotors, separated from the ground
state by gaps $\Delta_{\rm A}(S) \propto S(S+1)/L^2$, while the singlet excited state should be the gapped amplitude (``Higgs'') mode 
in the AF state \cite{Lee19}. In contrast, in other putative phases adjacent to the AF phase (in the SS model and many other models), the $S=2$ state 
will be above the lowest $S=0$ excitation. Thus, we identify the extrapolated singlet-quintuplet crossing point $g_{c2} \approx 0.82$ with a quantum 
phase transition into the AF state.

Following previous work on the $J_1$-$J_2$ model \cite{Wang18}, we identify the extrapolated singlet-triplet crossing point $g_{c1} \approx 0.79$ 
with the transition out of the PS state. The singlet minimum by itself is consistent with the PS gap vanishing at a DQCP and becoming the gapped
amplitude mode in the AF phase \cite{Lee19}. However, an AF phase starting at $g_{c1}$ is inconsistent with the singlet-quintuplet
crossing point $g_{c2}$. Though the separation between the transition points $g_{c1} \approx 0.79$ and $g_{c2} \approx 0.82$ is small, an 
eventual flow toward a common point for larger systems appears unlikely, given the absence of significant corrections to the $1/L^2$ forms 
in Fig.~\ref{gcfits}. Below we will show evidence for a gapless SL phase for $g \in (g_{c1},g_{c2})$.

Both gap crossings match those in the $J_1$-$J_2$ Heisenberg model \cite{Wang18}, 
where several numerical studies have reached a consensus on the existence of a gapless SL phase between 
dimerized and AF phases \cite{Gong14,Morita15,Wang18,Nomura20,Ferrari20}.  Field theories have also recently been proposed for this SL phase 
\cite{Schackleton21,Liu20}. Moreover, the same level crossings were found at the transition from a critical state 
to either a dimerized state (singlet-triplet crossing) or an AF state (singlet-quintuplet crossing) in a frustrated Heisenberg chain with 
long-range interactions \cite{Sandvik10,Wang18}. Given these results for related models, the distinct $g_{c1}$ and $g_{c2}$ points 
suggest a gapless SL phase also in the SS model.

\begin{figure}[t]
\includegraphics[width=65mm]{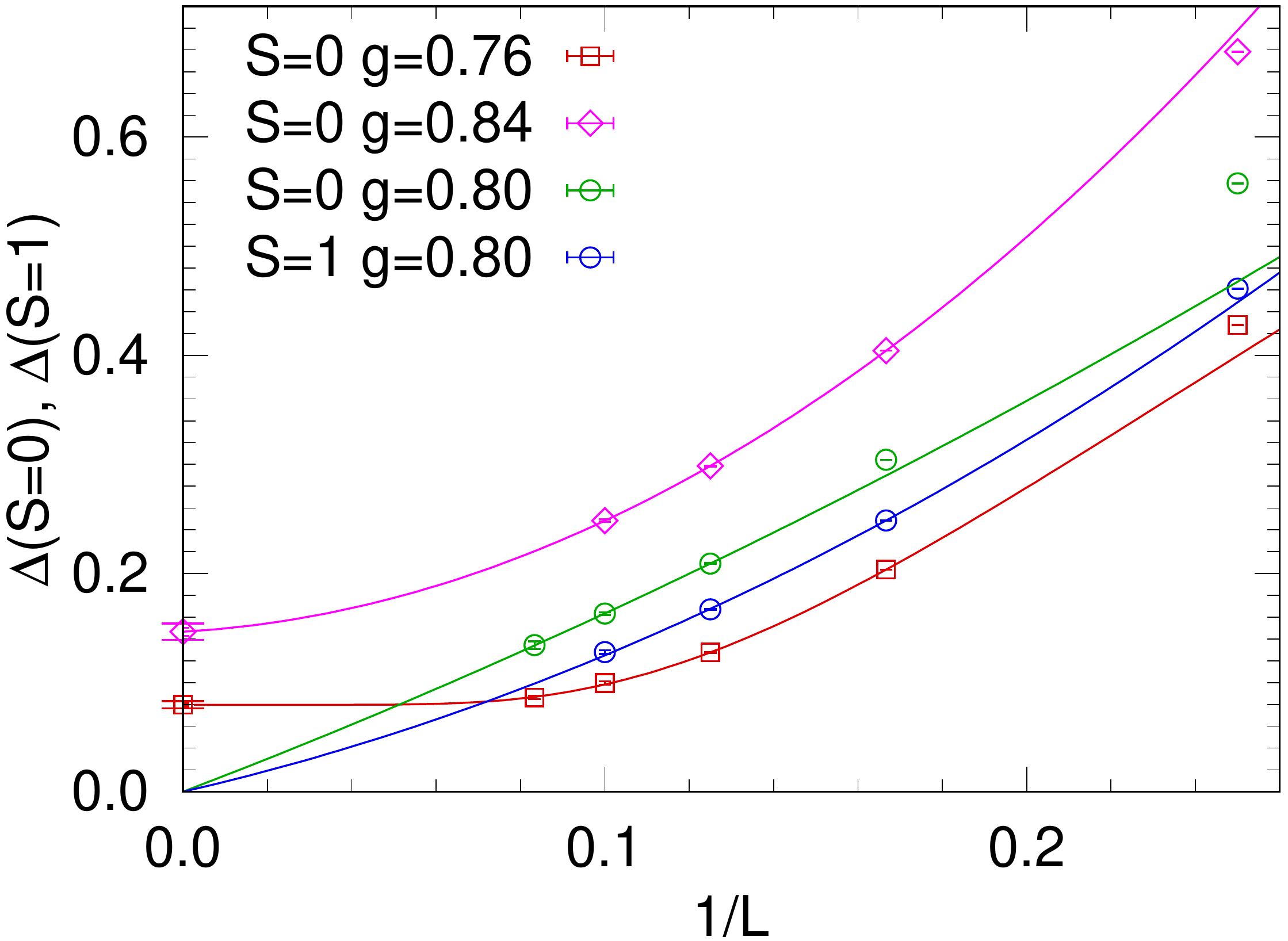}
\vskip-1mm
\caption{Gaps vs inverse system size. The singlet and triplet at $g=0.80$ (SL phase) have been fitted to the form $\Delta=a/L+b/L^2$
($a$ and $b$ being fitting parameters). The singlets in the PS ($g=0.76$) and AF ($g=0.84$) phases converge to non-zero values, as shown with 
a fit of the form $\Delta=a + b{\rm e}^{-cL}$ (fitting parameters $a,b,c$) in the former case and a quadratic form in the latter case.}
\label{gapscale}
\end{figure}

In Fig.~\ref{gapscale} we analyze the size dependent gaps in and close to the putative SL phase. At $g=0.80$, both the singlet and triplet 
gaps exhibit asymptotic $1/L$ scaling, corresponding to a dynamic exponent $z=1$ inside the SL phase. At $g=0.76$, in the PS 
phase, the singlet (and also the not shown triplet) converges exponentially to a non-zero gap, as expected in the SS model with 
cylindrical boundaries (Fig.~\ref{lattice}) for which the shifted PS state is gapped by boundary energies. In the AF phase, we find 
convergence to a non-zero amplitude-mode energy at $g=0.84$. In Fig.~\ref{gapscale} we have fitted a polynomial in this case, which works 
better than an exponentially convergent form, likely due to a gapless spectrum above the lowest singlet (unlike the isolated singlet 
mode in the PS state).

\begin{figure}[t]
\includegraphics[width=80mm]{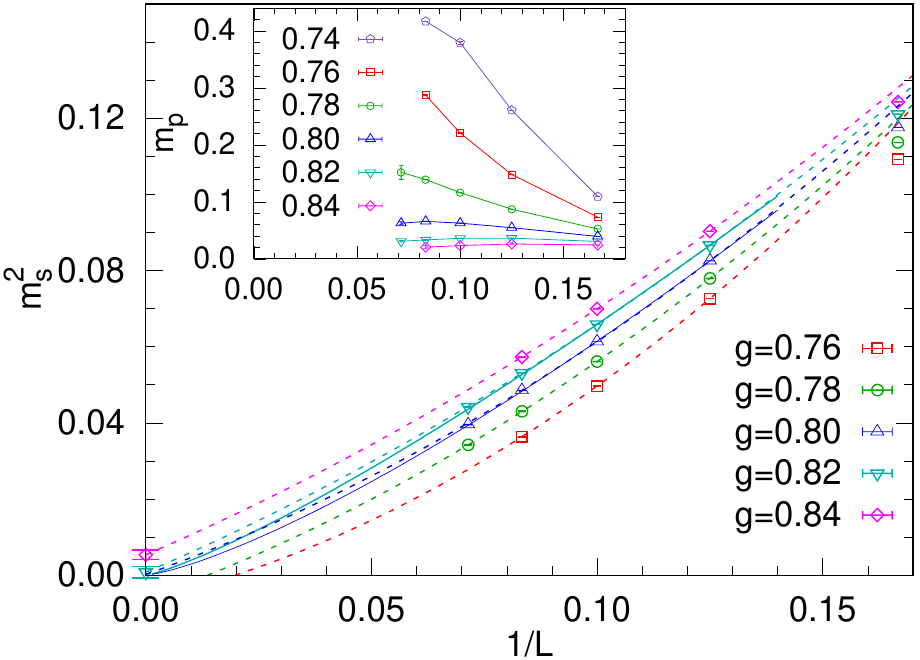}
\vskip-1mm
\caption{Squared AF order parameter vs inverse system size for several $g$ values. The corresponding PS order parameters are shown in the
inset. The dashed curves with colors matching the symbols in the main graph are second-order polynomials, while the solid curves are of the critical 
form $\propto L^{-(1+\eta)}$ with $\eta \approx 0.32$ and $\eta \approx 0.23$ for $g=0.80$ and $0.82$, respectively. Fitting to the $m_p$ data is not 
meaningful, but the non-monotonic behavior for $g=0.80$-$0.84$ is explained by boundary PS order outside the PS phase (Supplemental Material \cite{sm}) 
and $m_p \to 0$ for $L \to \infty$.}
\label{orderparam}
\end{figure}

We next study order parameters. We use the squared AF magnetization,
$m_s^2 = {L^{-4}} \sum_{ij}\phi_{ij}\langle \mathbf{S}_{i} \cdot \mathbf{S}_{j}\rangle$,
where $i,j$ are sites in the central $L\times L$
area of a $2L\times L$ system and $\phi_{ij} = \pm 1$ is the staggered phase. To detect PS order we define
$\mathbf{Q}_{\mathbf{r}}\equiv \frac{1}{2}(\mathbf{P}_{\mathbf{r}}+\mathbf{P}^{-1}_{\mathbf{r}})$, with $\mathbf{P}_{\mathbf{r}}$  a cyclic permutation
operator on the four spins of a plaquette at $\mathbf{r}$. Given the boundary-induced plaquette pattern (Fig.~\ref{lattice}), we can detect the PS
order as the difference of $\langle \mathbf{Q}_\mathbf{r}\rangle $ on two adjacent ``empty'' SS plaquettes \cite{Zhao20a}. Thus, we define
$m_p = \langle \mathbf{Q}_\mathbf{R}-\mathbf{Q}_\mathbf{R'}\rangle$, where $\mathbf{R}$ and $\mathbf{R'}$ are both close to the center of the cylinder
(the landscape of $\mathbf{Q}_\mathbf{r}$ values is shown in the Supplemental Material \cite{sm}). Both order parameters
are graphed versus $1/L$ in Fig.~\ref{orderparam}.

Second-order polynomial extrapolations of the AF order parameter in Fig.~\ref{orderparam} show that $m^2_s$ vanishes for $g \approx 0.82$, thus providing 
further evidence for the AF phase starting at the extrapolated singlet-quintuplet point $g_{c2} \approx 0.82$. The polynomial form is strictly appropriate 
only inside the AF phase, while at a critical point (or phase) $m^2_s \propto L^{-(1+\eta)}$ should instead apply asymptotically. The $g=0.80$ and $0.82$ 
data can indeed be fitted with $\eta \approx 0.32$ and $\eta\approx 0.23$, respectively. In the PS phase, polynomial fits extrapolate to unphysical 
negative values, which can be understood on account of the expected $\propto L^{-2}$ asymptotic form (which, however, cannot be fitted
because of large corrections).

The inset of Fig.~\ref{orderparam} shows how PS order is stabilized only for the larger system sizes inside the PS phase, reflecting large 
fluctuations in small systems (as shown explicitly in Supplemental Material \cite{sm}). The central plaquettes where $m_p$ is defined are close 
to the cylinder edges for small $L$, and only for larger $L$ can $m_p$ reflect a disordered bulk. Outside the PS phase the boundary-induced 
order close to the edges first increases with $L$, thus causing non-monotonic behavior as seen most clearly at $g=0.82$ and $0.84$ (see also 
Supplementary Material \cite{sm}). At $g=0.80$, $m_p$ for $L=14$ also falls below the value for $L=12$, indicating that indeed $m_p \to 0$ 
when $L \to \infty$, as it should in the SL phase.

{\it DQCP and unified phase diagram.}---The originally proposed DQCP is generic, reachable by tuning
a single parameter \cite{Senthil04}. Quantum Monte Carlo studies of several variants of $J$-$Q$ Hamiltonians \cite{Sandvik07}
have indeed found direct transitions between AF and dimerized ground states 
\cite{Melko08,Jiang08,Lou09,Sandvik10b,Kaul11,Sandvik12,Block13,Harada13,Chen13,Pujari15,Shao16,Sandvik20}. Similar results have been obtained
with related classical loop \cite{Nahum15a,Nahum15b} and dimer \cite{Sreejith19} models. In most cases, no discontinuities were observed, 
though unusual scaling violations point to weak first-order transitions \cite{Jiang08,Chen13,Wang21} or other scenarios
\cite{Sandvik12,Shao16}. One proposal is that the DQCP is unreachable (e.g., existing only in dimensionality below $2+1$)
and described by a nonunitary conformal field theory (CFT) \cite{Wang17,Gorbenko18a,Gorbenko18b,Ma20,Nahum20,Li18,He20}. 

In some variants of the $J$-$Q$ model clearly first-order transitions were observed \cite{Zhao19,Zhao20b,Takahashi20}. The Checker-Board $J$-$Q$
(CBJQ) model  \cite{Zhao19} (and a closely related loop model \cite{Serna19}) has a $\mathbb{Z}_2$ breaking PS phase like that in the SS model. A 
first-order spin-flop-like
transition with emergent O(4) symmetry of the combined O(3) AF and scalar PS order parameters was found, with no conventional coexistence
state with tunneling barriers up to the largest length scales studied. This unusual behavior indicates close proximity to an O(4) DQCP.

Lee et al.~recently considered a proxy of the excitation gap with the IDMRG method (infinite-size DMRG, where $L_x \to \infty$ and $L_y$ is finite), 
studying correlation lengths of operators with the symmetries of
the excited levels of interest \cite{Lee19}. Following Ref.~\onlinecite{Wang18}, they identified both crossing points discussed here (Figs.~\ref{stgaps}
and \ref{gcfits}), but these points were not extrapolated to infinite size. It was nevertheless argued that the singlet-triplet and singlet-quintuplet
crossings will drift to a common DQCP with increasing system size, in the SS model as well as in the $J_1$-$J_2$ model. However, in a very recent work,
Shackleton et al.~revisited the $J_1$-$J_2$ model and constructed a quantum field theory of a gapless SL phase and a DQCP separating it from
the AF state \cite{Schackleton21}. A different field theory was outlined in Ref.~\onlinecite{Liu20}.

\begin{figure}[t]
\includegraphics[width=55mm]{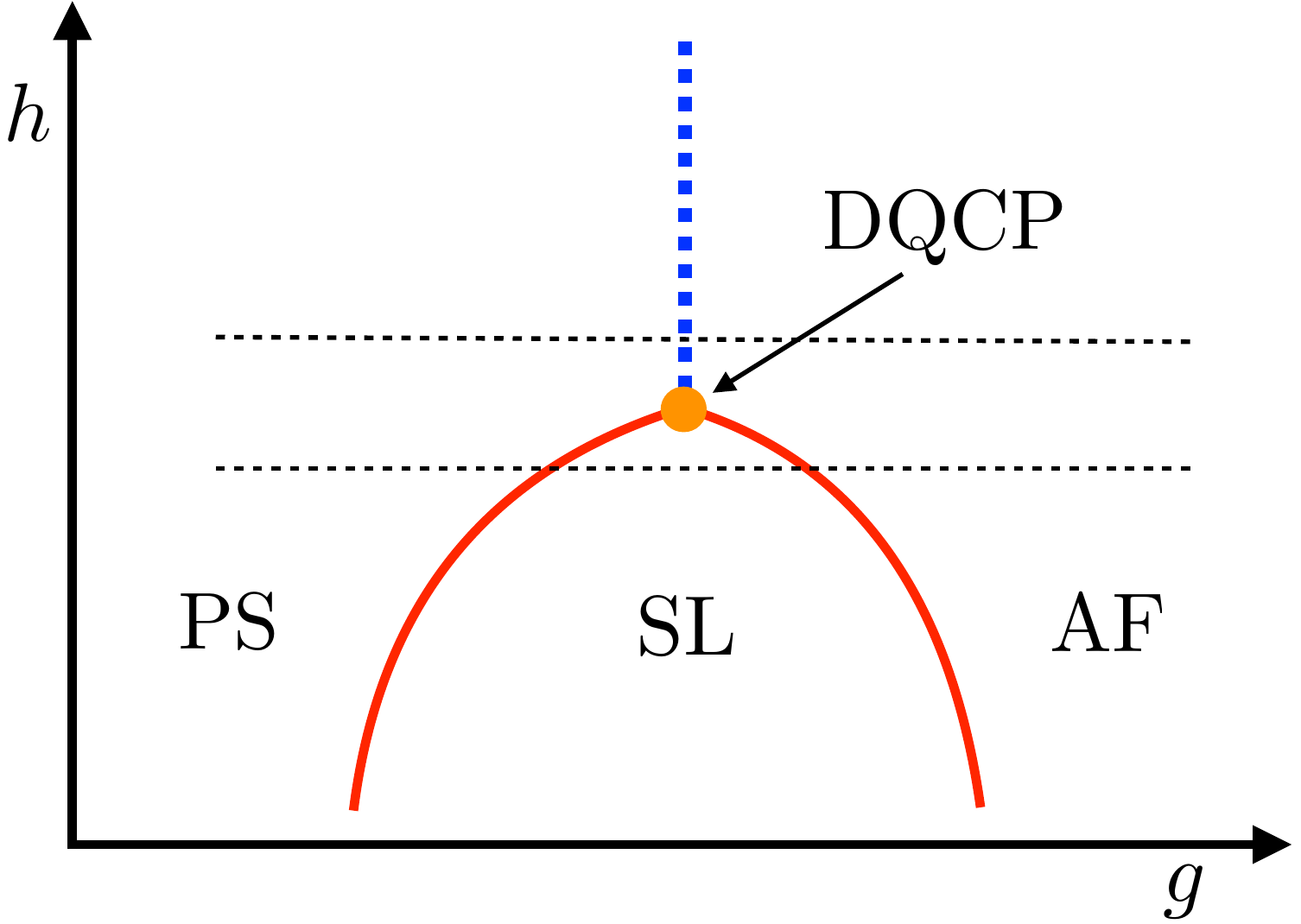}
\vskip-1mm
\caption{Unified phase diagram, where an O$(4)$ DQCP separates a line of first-order PS--AF transitions and an extended SL phase. The PS--SL and SL--AF
transition may both be continuous DQCP-like transitions. The dashed horizontal lines illustrate cuts through the phase diagram when a single parameter
$g$ is tuned; corresponding the CBJQ model (top line) and the SS model (bottom line).}
\label{phases}
\end{figure}

The narrow SL phase found here in the SS model suggests proximity to the DQCP discussed by Lee et al \cite{Lee19}, which most likely 
would be the same DQCP as the one influencing the O(4) transition in the CBJQ model \cite{Zhao19}. Moreover, it has recently been argued that 
the DQCP is actually a multi-critical point \cite{Zhao20c,Lu21}; a second relevant scaling field with all the symmetries of the Hamiltonian 
was detected in the conventional critical $J$-$Q$ model \cite{Zhao20c}, and subsequently such a field was also identified in a deformed 
self-dual field theory \cite{Lu21}.  In the $J$-$Q$ model it was found that the system flows toward a first-order transition when a certain 
interaction is turned on in a way maintaining a sign-free path integral \cite{Zhao20b}. It is possible that the interaction with the opposite 
sign could instead open up an SL phase. Taken together, all these observations suggest the unified phase diagram schematically illustrated 
in Fig.~\ref{phases}. The two parameters $(g,h)$ correspond to two relevant symmetric fields, and in models with just one tuning parameter $g$, 
e.g., the CBJQ and SS models, either the first-order line or the SL phase is traversed. Possible ways to tune $h$ in a model are 
further discussed below.

{\it Summary and Discussion.}---Our DMRG results can consistently be explained by a previously not anticipated SL phase between the known PS and
AF phases of the SS model. The PS--SL point $g_{c1} \approx 0.79$ is above the PS--AF point $g_{c} \approx 0.765$ obtained with tensor product states
\cite{Corboz13} (where the system is infinite but the results may be affected by small tensors) but is not at significant variance with the more recent
IDMRG calculation \cite{Lee19}, where $g_c \approx 0.77$ for $L=12$ and an increase in $g_c$ with $L$
was  observed (see Table 1 of Ref.~\onlinecite{Lee19}). The tensor technique used in Ref.~\onlinecite{Corboz13} has a bias to ordered phases and may
induce AF order in the fragile SL phase. In Ref.~\onlinecite{Lee19} the AF order parameter was not studied, and its appearance only at higher $g$
may have been missed. While these works did not consider any other phase intervening between the PS and AF phases, an early
field theory of the SS model within an $1/S_i$ expansion (with $S_i=1/2$ being the target spin value) contains phases not detected numerically
to date, including a gapped SL and a helical phase, but no gapless SL \cite{Chung01}. Topological order has also been proposed \cite{Ron14}.
As discussed further in the Supplemental Material \cite{sm}, for all values of $g$ we find the dominant spin correlations at the N\'eel 
wave-vector ${\bf k}=(\pi,\pi)$, i.e., no helical order.

Given our SS results and the existence of a gapless SL in the square-lattice $J_1$-$J_2$ model \cite{Gong14,Morita15,Wang18,Nomura20,Ferrari20,Schackleton21}, SLs ending at DQCPs may be ubiqutous between symmetry-breaking singlet and AF phases. The commonly studied Dirac SLs should be unstable on square lattices and 
lead to DQCPs \cite{Lee19,Song20}, and the SL identified here should fall outside this framework \cite{Schackleton21,Liu20}. In our scenario, in a
multi-parameter model the SL can be shrunk to a multi-critical DQCP with emergent symmetry, followed by a first-order direct PS--AF transition. In 
principle there  could be a triple point instead of the DQCP in Fig.~\ref{phases}, with weak first-order transitions as in the non-unitary CFT 
proposal \cite{Wang17,Gorbenko18a,Gorbenko18b,Ma20,Nahum20,He20}.

A DQCP separating a line of first-order transitions and an SL phase is a compelling scenario also considering that the $J$-$Q$ models can 
be continuously deformed into conventional frustrated models. Placing $Q$ terms on the empty plaquettes of the SS lattice, by gradually turning 
off $Q$ and turning on $J'$ the unusual first-order PS--AF transition with emergent O(4) symmetry of the CBJQ model \cite{Zhao19} should 
evolve as if the upper dashed line in Fig.~\ref{phases} moved to lower $h$, and eventually the SS SL phase should appear. In general, we 
expect that many perturbations of the SS and $J_1$-$J_2$ models could act as the parameter $h$ in Fig.~\ref{phases}, e.g., longer-range 
interactions or multi-spin cyclic exchange with appropriate signs. The O(4) symmetry should be replaced by SO(5) in cases where the PS phase
is instead a dimerized phase, e.g., with some extensions of the conventional J-Q and$J_1$-$J_2$ models.

An SL phase can explain the absence of any observed phase transition in SrCu$_2$(BO$_3$)$_2$ at pressures 2.6-3 GPa \cite{Guo20,Jimenez20}, 
between the PS and AF phases. Since SrCu$_2$(BO$_3$)$_2$ can be synthesized with very low concentration of impurities, unlike many other potential 
spin liquid materials, an SL phase would be significant. A direct PS-AF transition has already been observed at high magnetic fields 
\cite{Jimenez20}, but the nature of the transition remains unexplored. The phase diagram in Fig.~\ref{phases} may also hold with $h$ corresponding 
to the field strength, but with the symmetry of the AF order reduced to O$(2)$ and potentially emergent O$(3)$ symmetry of the DQCP [instead of 
O$(4)$ at zero field] and on the adjacent direct PS--AF line. 

{\it Note added.}---A recent functional renormalization group calculation, partially stimulated by our work, supports a gapless SL phase 
in roughly the same parameter regime as reported here \cite{Keles22}. Moreover, a study with tensor-product states of the J$_1$-J$_2$-J$_3$ 
Heisenberg model detected an isolated SL phase ending at a DQCP \cite{Liu21}, very similar to our phase diagram in Fig.~\ref{phases} 
when $g$ and $h$ are identified with $J_2/J_1$ and $J_3/J_1$, respectively ((and with a dimerized Z$_4$ phase instead of the Z$_2$ PS phase)
However, a line of continuous dimerized--AF transition was proposed beyond the SO(5) DQCP, instead of the weakly first-order transitions argued for here.

\begin{acknowledgments}
{\it Acknowledgments}.---We would like to thank Fr\'ed\'eric Mila, Zheng-Cheng Gu, 
Didier Poliblanc, Subir Sachdev, Sriram Shastry, Cenke Xu, and Yi-Zhuang You for stimulating 
discussions and comments, and Subir Sachdev also for sending us an early version of Ref.~\onlinecite{Schackleton21}.
A.W.S. was supported by the Simons Foundation under Simons Investigator Grant No.~511064.
L.W. was supported by the National Key Research and Development Program of China, Grant No.~2016YFA0300603,
and by the National Natural Science Foundation of China, Grants No.~NSFC-11874080 and No.~NSFC-11734002.
The computational results presented here were achieved partially using Tianhe-2JK computing time awarded
by the Beijing Computational Science Research Center (CSRC).
\end{acknowledgments}

\begin{widetext}
  
\newpage

\begin{center}  

{\large Supplemental Material}
\vskip3mm

{\bf\large Quantum Criticality and Spin Liquid Phase in the Shastry-Sutherland model}

\vskip3mm

Jianwei Yang,$^1$, Anders W. Sandvik,$^{2,3,*}$, Ling Wang$^{4,\dagger}$

\email{sandvik@bu.edu}
{\it 
{$^1$ Beijing Computational Science Research Center, 10 East Xibeiwang Road, Beijing 100193, China}\\
{$^2$ Department of Physics, Boston University, 590 Commonwealth Avenue, Boston, Massachusetts 02215, USA}\\
{$^3$ Beijing National Laboratory for Condensed Matter Physics \\ and Institute of Physics, Chinese Academy of Sciences, Beijing 100190, China}\\
{$^4$ Department of Physics, Zhejiang University, Hangzhou 310000, China}}
\vskip3mm
  
$^*$ e-mail: sandvik@bu.edu, $^\dagger$ lingwangqs@zju.edu.cn

\end{center}

\vskip3mm
Here we present additional results in support of the conclusions drawn in the main paper.
In Sec.~1 we discuss the DMRG procedures and illustrate the convergence properties and extrapolations of energies
and order parameters. In Sec.~2 we present 2D plots of the PS order ``landscape'' and discuss the effects of the open $x$
boundaries of the cylindrical lattices, including an explanation for the nonmonotonic behavior of the PS order parameter
for $g \ge 0.80$ in Fig.~\ref{orderparam}. In Sec.~3 we present the static spin structure factor in the full wave-vector space,
which shows no evidence of helical magnetic correlations in the SS model.
\vskip6mm

\end{widetext}

\setcounter{page}{1}
\setcounter{equation}{0}
\setcounter{figure}{0}
\renewcommand{\theequation}{S\arabic{equation}}
\renewcommand{\thefigure}{S\arabic{figure}}
\renewcommand{\thesection}{\arabic{section}}

\subsection*{1. DMRG convergence and extrapolations}

In order to systematically approach the correct ground states and excitations in the DMRG calculations,
we carry out calculations for several numbers $m$ of Schmidt states, until either $m$ is sufficiently large
for the discarded weight $\epsilon$ to be negligible (in case of the smaller system sizes considered here) or
sufficiently small for reliably extrapolating to $\epsilon=0$ (for the largest system sizes). The discarded
weight is defined in the standard way as the sum of discarded eigenvalues of the reduced density matrix
\cite{Schollwock11,Stoudenmire13}. 

After an initial calculation for small $m$, each subsequent calculation is started from either a previously well converged calculation 
for a smaller $m$ and the same $g$ value, or one for a nearby $g$ value and the same $m$. For a given $m$, we demand that the
energy difference (not divided by the system volume) between two successive updating sweeps is less than
$10^{-6}$. We then check the convergence of the energies and other quantities as a function of the discarded
weight $\epsilon$ (which depends on $m$, with $\epsilon\to 0$ as $m\to \infty$).

\begin{figure*}[t]
\includegraphics[width=155mm]{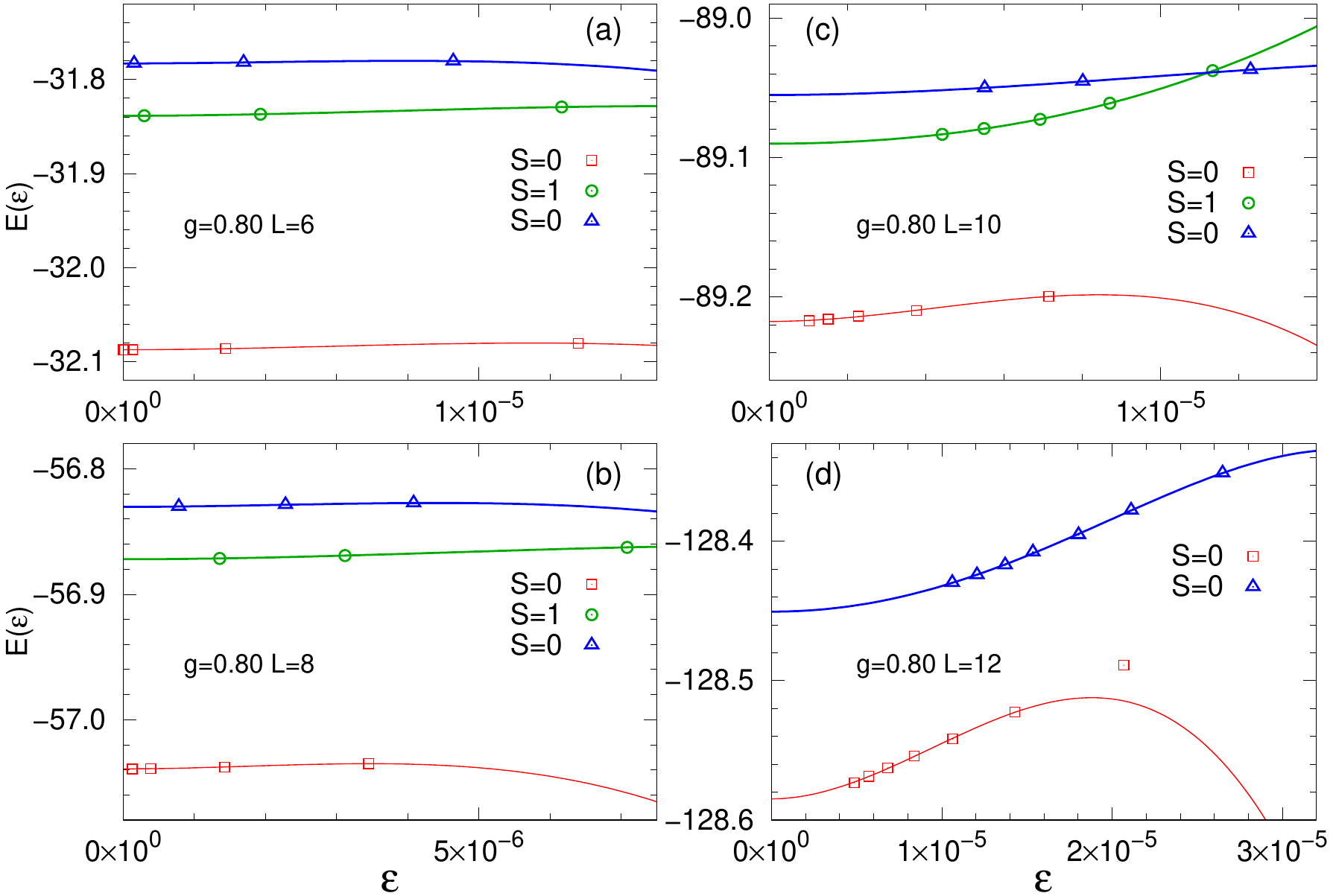}
\caption{Examples of extrapolations of DMRG energies to vanishing discarded weight $\epsilon$ for the SS model with $g=0.8$.
For each of the system sizes $L=6,8,10,12$ in panels (a)-(d), the two lowest singlets are shown (squares and triangles)
along with the lowest triplet (circles), except for $L=12$, for which we do not have sufficiently well converged triplet data. The
fitted curves are of the form $E(\epsilon)=E(0)+a\epsilon^2 + b\epsilon^3$, with $E(0)$, $a$, and $b$ optimized parameters. 
\label{gapextrap}}
\vskip5mm
\includegraphics[width=155mm]{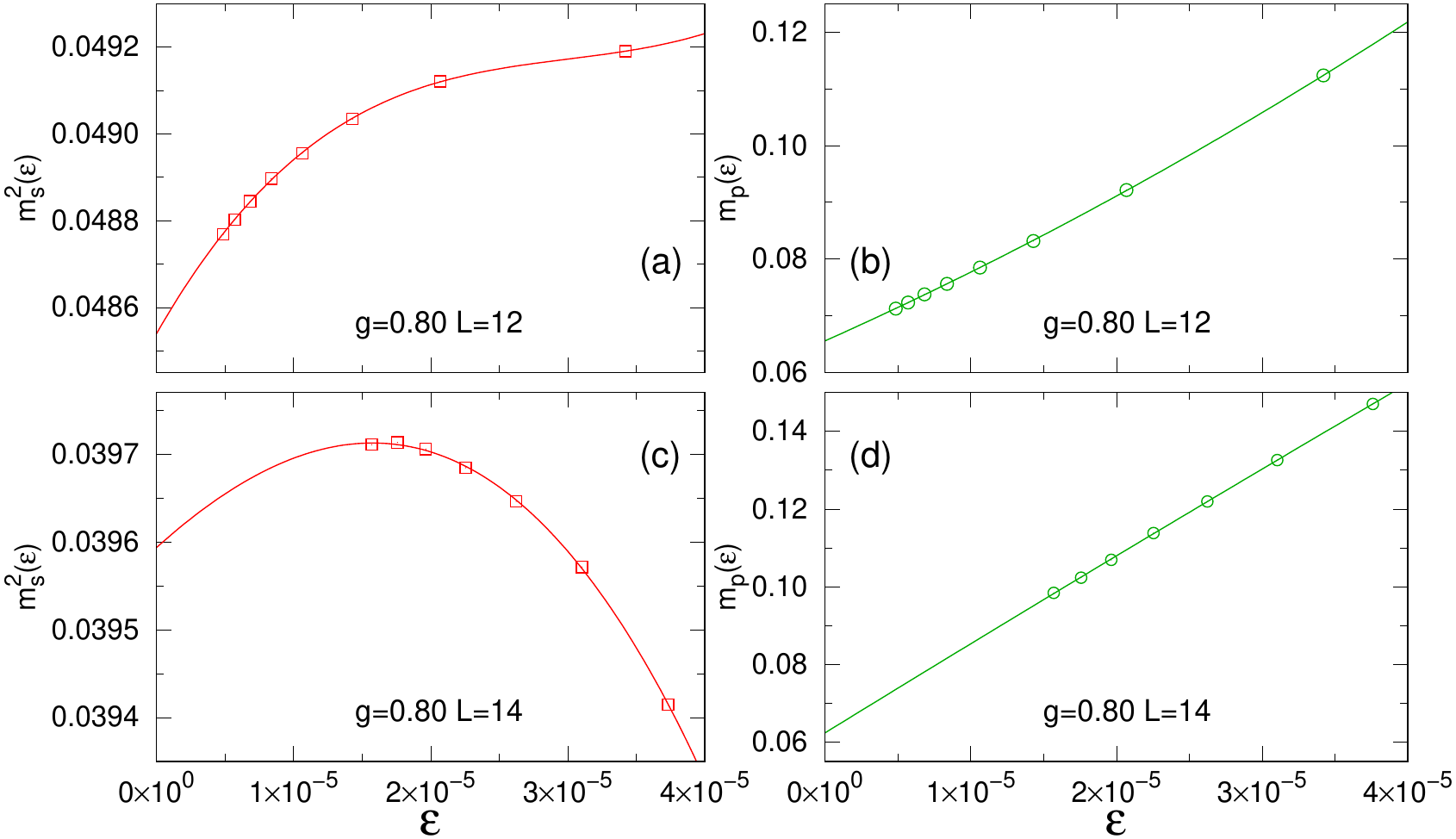}
\caption{Examples of extrapolations of DMRG results for the AF and PS order parameters in the ground state of the SS model at $g=0.8$.
Results for system size $L=12$ are shown in (a) and (b), and corresponding results for $L=14$ are shown in (c) and (d). The fitted curves
are quadratic polynimials in (a), (c) and cubic polynomials in (b),(d).}
\label{ordextrap}
\end{figure*}

Our DMRG program is implemented with full SU(2) symmetry \cite{Wbaum12,Gong14}, and we can therefore target the ground
state of any sector of total spin $S$. Given that the lowest ground state among these has $S=0$ and we are also interested
in the lowest $S=0$ excitation, in this sector we apply the technique of orthogonalizing to the previously calculated
ground state in order to target the lowest excited $S=0$ state \cite{McCulloch07,Wang18,Lemm20}.

\subsubsection{A. Cylindrical $2L \times L$ lattices}

In Fig.~\ref{gapextrap} we show the lowest two $S=0$ energies as well as the lowest $S=1$ energy for system sizes $L=6,8,10$,
and $12$, in the important case of $g=0.80$ (inside the new gapless SL phase). For the challenging $L=12$ system, we used
$m$ up to $9000$ for the $S=0$ ground state and up to $m=10000$ for the excited singlet. We do not have sufficiently good triplet
results for $L=12$ and therefore only show $S=1$ results for the smaller systems in Fig.~\ref{gapextrap}.

For $L=6$, the calculation for the largest $m$ already has an extremely small discarded weight, and it is not necessary to further extrapolate 
the results. We nevertheless show extrapolations for all system sizes in Fig.~\ref{gapextrap}. While the exact form of the error as a function 
of $\epsilon$ in the DMRG method is not known, in all cases for which data are presented in the main paper our results are sufficiently
converged for the extrapolated values to not be very sensitive to the fitting form used. We have found that the energies for
small $\epsilon$ are well described by third-order polynomials without linear term. Thus, in Fig.~\ref{gapextrap} all data are
fitted to such a form.

\begin{figure}[t]
\includegraphics[width=75mm]{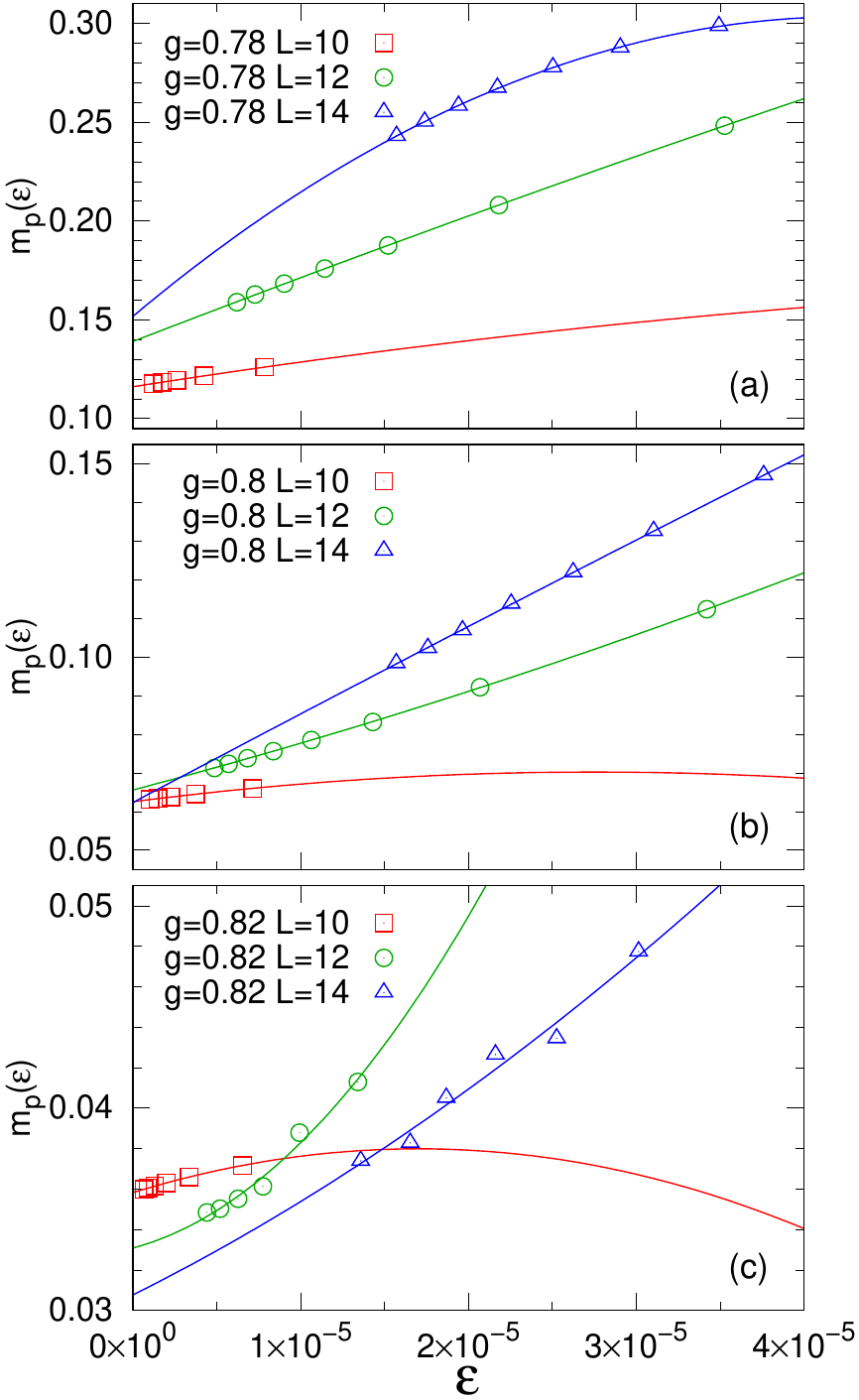}
\caption{Plaquette order parameter vs discarded weight for (a) $g=0.78$, (b) $g=0.80$, and (c) $g=0.82$. Each panel shows results
for system sizes $L=10,12,14$, and the curves are second-order polynomial fits used for extrapolating to $\epsilon=0$. The scatter of the $L=14$
data (and to a lesser extent for $L=12$) in (c) originates from incomplete convergence at the Lanczos stage of the DMRG calculation for some of 
the points. These arrors are reflected in the final error bars of the extrapolated $m_p$ value.}
\label{plqepsilon}
\end{figure}

In the case of the order parameters, we find that polynomial fits work well in general (here including the linear terms). We typically 
use quadratic or third-order forms. Examples at $g=0.8$ for the two largest system sizes, $L=12$ and $L=14$, are shown in Fig.~\ref{ordextrap}. 
Here we observe nonmonotonic behavior of the AF order parameter $m_s^2$ for $L=14$ in Fig.~\ref{ordextrap}(c). For the small $\epsilon$ values 
(large $m$) used in the extrapolations, we have found such nonmonotonic behavior of the largest systems for $g$ only outside the AF phase. 

We point out that the $m_s^2$ data points for $L=14$ in Fig.~\ref{ordextrap}(c) only exhibit nonmonotonicity in the two points for smallest $\epsilon$, 
but even when these points are excluded the fit still results in a nonmonotonic polynomial. Hence, we judge this behavior as stable and the
extrapolation as reliable. It should also be noted that the overall range of the $y$-axis of Fig.~\ref{ordextrap}(c) represents a vary small relative 
change in the value, and the uncertainties in the extrapolation for this case (less than $0.3\%$ between different extrapolations) do not impact 
the conclusions drawn on the basis of the data in Fig.~\ref{orderparam}.

As seen in Fig.~\ref{ordextrap}, the PS order parameter shows much larger overall dependence on $\epsilon$ than does the AF order parameter. An
important aspect of the PS order, discussed in the context of the inset of Fig.~\ref{orderparam} in the main paper, is that its size dependence is 
nonmonotonic outside the PS phase; in the SL phase as well as in the AF phase close to the SL transition. In Sec.~2 we will further illustrate how 
this behavior originates from the cylindrical boundary conditions. To further illustrate the nonmonotonic behavior and its robustness in the $\epsilon$ 
extrapolations, in Fig.~\ref{plqepsilon} we show $m_p$ results for the three largest system sizes, $L=10,12,14$, for three values of $g$ in and close 
to the SL phase. The nonmonotonic size dependence of the $\epsilon \to 0$ values is clear for $g=0.82$, and also for $g=0.80$ does this trend begin to 
appear. In the case of $g=0.78$, the extrapolated order parameter only grows with $L$. 

Based on only these data, it is of course not possible to exclude a maximum of the $\epsilon \to 0$ values  versus $L$ followed by an eventually 
decrease to $0$ for larger system sizes also 
for $g=0.78$. However, the other results in the main paper show that this $g$ value is inside the PS phase, and, therefore, $m_p$ should flatten out 
to take a finite value if $L$ is further increased (as is seen for still smaller $g$ values in Fig.~\ref{orderparam} in the main paper). The observed 
nonmonotonic behavior outside the PS phase (with its physical explanation in Sec.~2 below) allows us to put an upper bound on the extent of the PS phase, 
$g_{c1} < 0.80$, which is consistent with the SL phase boundaries in Fig.~\ref{gcfits} of the main paper.

\begin{figure*}[t]
\includegraphics[width=17cm]{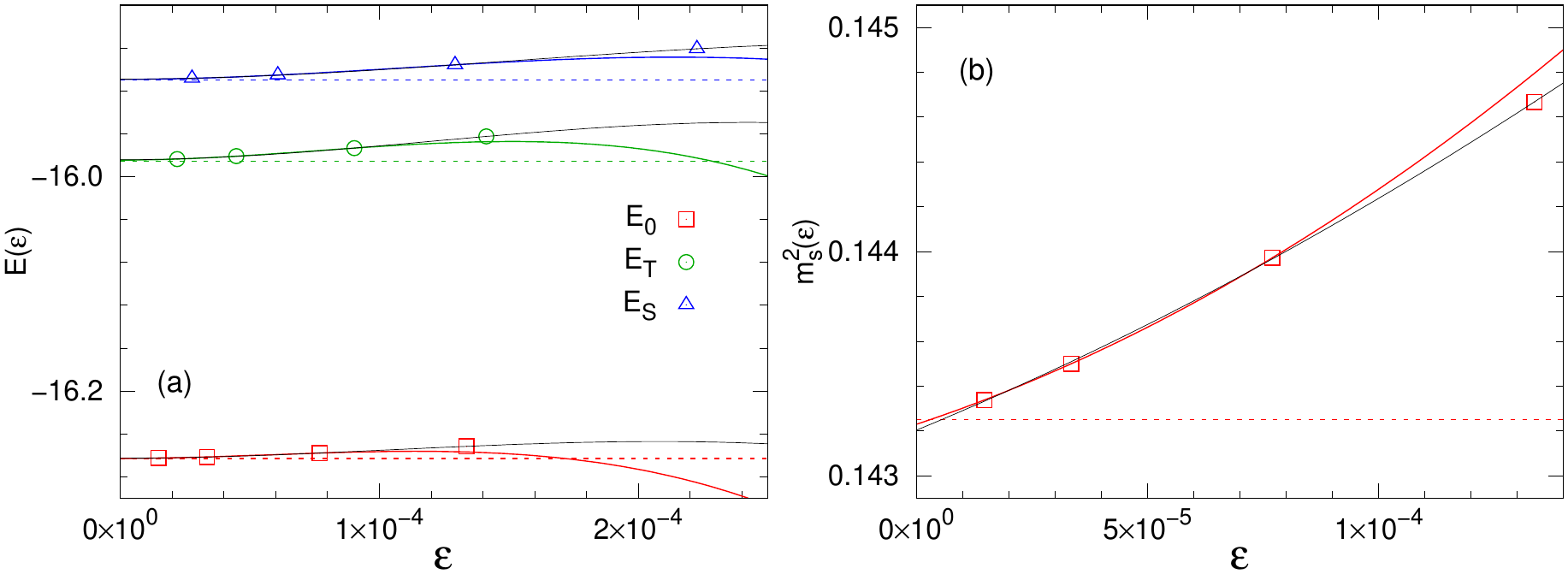}
\vskip-1mm
\caption{Exact and DMRG results for the $6\times 6$ periodic SS lattice at $g=0.80$. (a) DMRG energies of the ground state and first 
excited state in the singlet and triplet sectors vs the discarded weight $\epsilon$, obtained with bond dimension $m=800$, $1200$, 
$2000$, and $3000$. The colored dashed lines are the corresponding eigenvalues obtained by Lanczos exact diagonalization. The colored solid 
curves show fits to the  form $E=a+b\epsilon^2+c\epsilon^3$ using the three points with smallest $\epsilon$. The black solid curves show fits to the same 
functional form using all four points. Their difference in the $\epsilon \to 0$ extrapolated values defines the DMRG error of the energy.
(b) The staggered magnetization of the ground state. Colored (black) solid curves show fits to the quadratic functions $m_s^2=a+b\epsilon+c\epsilon^2$ 
using three (four) points, and their extrapolation difference defines the DMRG error of the staggered magnetization. 
The red dashed line shows the exact value.}
\label{enr_extr}
\end{figure*}

\subsubsection{B. $6\times 6$ periodic system}

To further illustrate the soundness of the extrapolation procedures and our methods of estimating the errors of the $\epsilon \to 0$ 
energies and order parameters, we next consider a $6\times 6$ lattice with fully periodic (toroidal) 
boundary condition. Its eigenenergies for the ground state, the first singlet and triplet
excitations, as well as the ground state order parameters, were obtained by Lanczos exact diagonalization. Using DMRG 
with SU(2) symmetry, we computed the corresponding set of variational energies and order parameters as a function of 
discarded weight $\epsilon$ using several bond dimensions $m$. Even though the Hilbert space here is small
enough for Lanczos diagonalization, the convergence properties for moderate values of $m$ are very
similar to those discussed above for the larger cylindrical lattices. Extrapolations as performed 
above can therefore serve as a bench-mark for the reliability of the procedures.

Figure \ref{enr_extr}(a) shows variational energies for the $6\times 6$ SS lattice at $g=0.80$ obtained by SU(2) DMRG with bond dimensions 
$m=800$, $1200$, $2000$, adn $3000$. The fitting function is a third-order polynomial, again without the linear term ($E=a+b\epsilon^2+c\epsilon^3$). 
For a reasonable definition of the error of extrapolation to $\epsilon\to 0$, we take the difference between extrapolated 
energies based on the largest three $m$ values, shown with colored curves in Fig.~\ref{enr_extr}(a), and all four points,
shown with in black curves. The colored dashed horizontal lines mark the exact energies, which are $E_0=-16.263112,
E_{\rm T}=-15.985471, E_{\rm S}=-15.910097$, for the ground state, singlet excitation, and triplet excitation, respectively.
The DMRG extrapolation errors defined as above are $0.00017$, $0.0003$, and $0.0003$, respectively. 

The ``exact error'', 
which we for the sake of the illustration here (where we know the exact energies, which is of course not of case for the
cylindrical lattices used in our primary studies of the SS model) define as the difference between extrapolated value based 
on three data points and the exact energy, are $0.00022$, $0.0008$, $0.0007$, respectively. If the fitting function is switched to 
a second order polynomial (including the linear term), the extrapolated results using all four $m$ points go slightly below the exact 
energies for all three states.  Overall, we again found that the higher-order polynomials without the linear terms describe 
the data better.

The above example shows that the estimated extrapolation errors are of the same magnitude as the ``exact errors''.
Here the minimum available discarded weight, corresponding to $m=3000$ is $\epsilon\approx 2\times 10^{-5}$. This $\epsilon$ 
is much larger than the truncation errors reached with larger $m$ in cylinder systems for $L\leq 10$ and is comparable with 
the $\epsilon$ value reached for $L\geq 12$, as seen in Fig.~\ref{gapextrap}. The way the energies flatten out for the cylindrical 
systems when $\epsilon$ decreases is very similar to what we observe for the $6 \times 6$ system, and we see no reason why the 
results should not have reached $\epsilon$ small enough to perform the extrapolations as explained.

Figure \ref{enr_extr}(b) shows extrapolations of the staggered magnetization $m_s^2(\epsilon)$ in the ground state for which the energies are 
shown in Fig.~\ref{enr_extr}(a). Here the fitting function is a regular second-order polynomial, $m_s^2=a+b\epsilon+c\epsilon^2$. The error of 
the order parameter is again defined as the difference between extrapolated values using either the three largest $m$ points or all four points. 
The exact value from the Lanczos exact diagonalization is $m_s^2=0.143250$. The error of the DMRG extrapolation is $0.000026$ in this case, and the 
``exact error'' as defined above is $0.000021$. We again see that the defined extrapolation error is very reasonable. 

All finite-size errors of the results discussed in the main paper were estimated as above, but in most cases the error bars are much smaller than the
graph symbols and are not shown explicitly. In the extrapolations to infinite size, the finite-size errors were propagated in the standard way in the
function fits, and the resulting error bars are displayed in some of the figures.

\begin{figure*}[t]
\null\hskip-8mm\null\includegraphics[width=185mm]{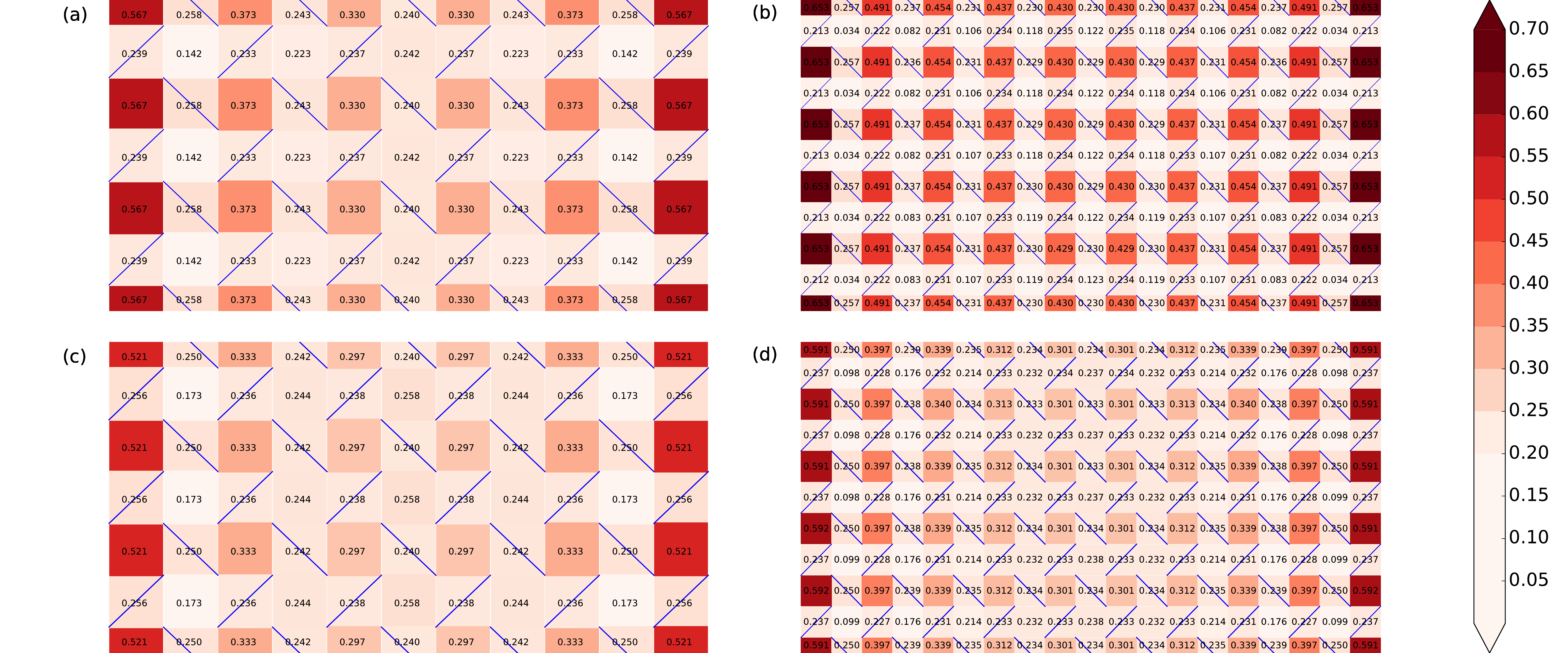}
\caption{Landscape of plaquette singlet strengths in systems of size $L=6$ (left) and $L=10$ (right), for $g=0.75$ (inside the PS phase) in (a) and 
(b), and for $g=0.80$ (in the SL phase) in (c) and (d). The colored squares correspond to the expectation value $\langle \mathbf{Q}_\mathbf{R} \rangle$
of the plaquette operator defined in the main paper for the plaquettes at lattice coordinate $\mathbf{R}$. The SS dimer couplings $J'$ are 
indicated with the blue lines. In addition to the color coding shown on the vertical bar, the actual numerical values of
$\langle \mathbf{Q}_\mathbf{R} \rangle$ are also displayed inside each square.}
\label{pslandscape}
\end{figure*}

\subsection*{2. Role of cylinder edges}

As shown in Fig.~\ref{lattice} in the main paper, the PS ordering pattern is unique on the cylindrical lattices used here. The cylinder edges act
as a $\mathbb{Z}_2$ symmetry-breaking ``pinning field'', allowing us to study the PS order parameter $m_p$ directly, instead of using the squared order 
parameter in a system with unbroken symmetry. This approach was discussed in detail in Ref.~\onlinecite{Zhao20a} in the context of a different system, and
it was argued that it is the best way to study the order parameters of ``singletized'' phases with methods that use symmetry-breaking boundary conditions.

Fig.~\ref{pslandscape} shows examples of the singlet pattern forming on two different lattices, of size $L=6$ and $L=10$, both inside the PS
phase at $g=0.75$ and in the SL phase at $g=0.80$. In the PS phase, we can observe that the alternating pattern of strong and weak empty
plaquettes (those without the SS diagonal couplings $J'$) is much stronger in the larger systems. This order enhancement with increasing $L$ in
the PS phase was already seen in the size dependent $m_p$, defined as the difference between central adjacent empty plaquettes, in the inset of
Fig.~\ref{orderparam} in the main paper. The strengthening of the PS order clearly reflects the diminishing quantum fluctuations when long-range
order is established with increasing system size in the presence of the symmetry-breaking edge field. Note that the edge order also strengthens 
with increasing $L$.

Turning now to the results in the SL phase, Fig.~\ref{pslandscape}(c) and \ref{pslandscape}(d), here as well we observe how the edge order is significantly
stronger in the larger system. The bulk order parameter, defined in the center of the system, is also stronger in the larger system. However, as seen in
the inset of Fig.~\ref{orderparam} in the main paper, for the largest system size considered for $g=0.80$, $L=14$, $m_p$ has turned downward [and
this is even more clear at $g=0.82$ as also illustrated by the results in Fig.~\ref{plqepsilon}(c)]. This 
nonmonotonic behavior outside the PS phase can naturally be explained as a competition between the always present (for any $g$) symmetry breaking at the 
cylinder edge and the decay of this ``artificial'' induced order in the central part of the system as $L$ increases. The initial increase with $L$ for small systems 
is due to the strengthening of the edge order with $L$ even when the system is in the SL or AF phase. The eventual down-turn of the order parameter for 
larger systems is a sign of this edge effect not extending to the bulk, i.e., that the system is not in the PS phase.

It is difficult to imagine any realistic mechanism that would cause $m_p$ to turn back up as $L$ increases further after the peak value has been 
reached and $m_p$ has begun to decrease with $L$. Therefore, we regard the observation of a maximum
in $m_p$ for a given $L$ as a definite indicator of the system not being in the PS phase. Our results for $g=0.80$-$0.84$ in Fig.~\ref{orderparam}
all exemplify this behavior.

\subsection*{3. Spin structure factor}

\begin{figure*}[t]
\null\includegraphics[width=170mm]{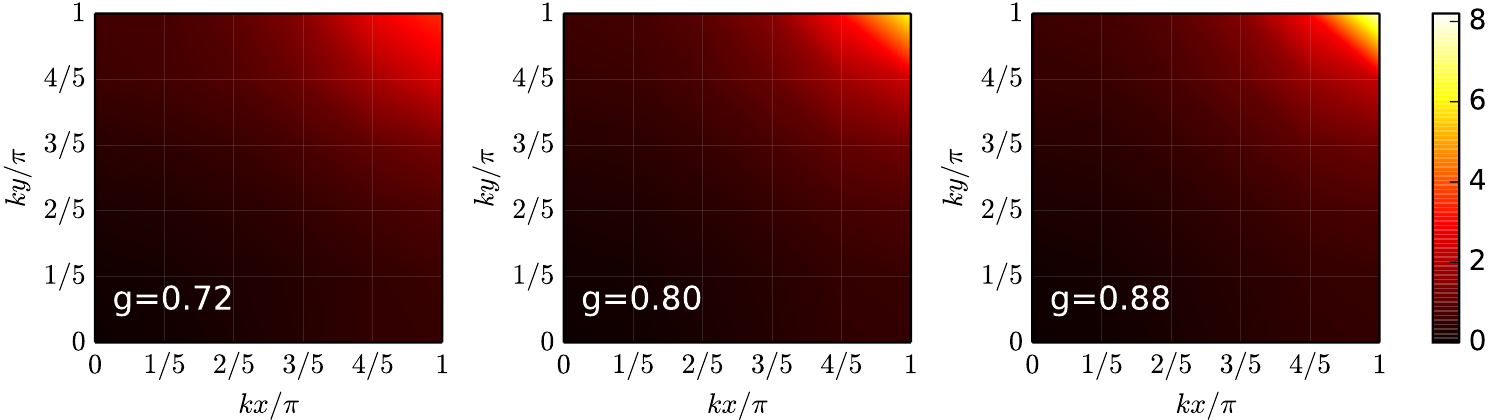}
\caption{The spin structure defined in Eq.~(\ref{sk}) for $L=10$ SS lattices at three different $g$ values as indicated. The slightly broken 
$k_x \leftrightarrow k_y$ reflection symmetry is due to the cylindrical $N=2L \times L$ lattice geometry, that is manifested even when the structure 
is factor defined only on the central $L \times L$ sites according to Eq.~(\ref{sk}). The symmetry is gradually restored for increasing $L$.}
\label{structure}
\end{figure*}

In an early field-theoretical study of the SS model within the framework of $1/S_i$ expansions ($S_i$ being the spin quantum number of the individual spins)
\cite{Chung01}, several exotic phases were found in the plane $(g,1/S_i)$, including a gapped SL and a phase with helical magnetic order. The gapped SL is
located close to a PS state but not between the PS and AF states as we have found here in the case of the gapless SL. The AF, PS, SL, and helical phases
meet at a point, which should be contrasted with our proposed unified phase diagram for DQCPs adjacent to gapless SL phase in Fig.~\ref{phases}, where the
PS, AF, and gapless SL phases meet at a  DQCP (or possibly a triple point of first-order transitions).

The SL phase of Ref.~\onlinecite{Chung01} also hosts short-range helical spin order. The proposed helical phase and the helical correlations of the gapped SL phase
prompts us to investigate the possibility of helical spin correlations in the gapless SL phase identified in the phase diagram of the SS model in the present
work. To this end, we have calculated the static spin structure factor $S({\bf k})$ based on the spin correlation functions
$\langle {\bf S}_i \cdot {\bf S}_i\rangle$ for all spins $i,j$ within the central $L \times L$ square of sites on the cylindrical $2L \times L$ lattices, defining
\begin{equation}
S({\bf k}) = \frac{1}{L^2} \sum_{i,j} {\rm e}^{-i{\bf k} \cdot ({\bf r}_i-{\bf r}_i)}  \langle {\bf S}_i \cdot {\bf S}_i\rangle. 
\label{sk}
\end{equation}
In a periodic $L \times L$ system the wave-vectors should take the form $k_a = n_a2\pi/L$, for $a \in \{x,y\}$ and $n_a \in \{0,\ldots,L-1\}$.
However, since we do not have periodic boundaries in the $x$ direction, we can evaluate the Fourier transform for any $k_x \in [0,2\pi]$. To obtain a smooth 
representation of the structure factor in the 2D $k$ space, we also use a large number of values for $k_y$ between the discrete points in principle allowed 
by the lattice geometry. Because of the periodicity in $k$-space we limit the values to $k_x,k_y \in [0,\pi]$.

In Fig.~\ref{structure} we show results for $L=10$ systems at three values of the coupling ratio, representing the PS, ($g=0.72$), SL ($g=0.80$), and
AF ($g=0.88$) phases. In all cases we observe that the structure factor is peaked at ${\bf k}=(\pi,\pi)$ indicating that the correlations are of the
N\'eel AF type in all phases. 

Given the rather small lattice size, we can of course not rigorously rule out order or dominant short-range correlations at
${\bf k}=(\pi,q)$ or ${\bf k}=(q,\pi)$ (where the cylindrical lattices would favor one case) with $q$ very close to $\pi$, but we find such a
scenario unlikely.

\end{document}